\documentclass{raa}            

\usepackage{graphicx,times}             
\usepackage{natbib}
\usepackage{amssymb,amsmath}
\usepackage{CJK}
\usepackage{xcolor}
\usepackage{soul}
\usepackage[pagebackref=true]{hyperref}
\begin{document}
\begin{CJK*}{UTF8}{gbsn}
  \title{Discovery of extended structure around open cluster COIN-Gaia 13 based on Gaia EDR3
  
  }

   \volnopage{Vol.0 (20xx) No.0, 000--000}      
   \setcounter{page}{1}          
   
   \author{Leya Bai (白乐娅) 
      \inst{1,2,3}
   \and Jing Zhong (钟靖）
      \inst{2}
      \footnote{ Corresponding authors：Jing Zhong \& Jing Li }
   \and  Li Chen (陈力)
      \inst{2,3}
   \and  Jing Li (李静)
      \inst{1}
    \and  Jinliang Hou (侯金良)
      \inst{2,3}
   }
   \institute{\textbf{School of Physics and Astronomy},China West Normal University, 1 Shida Road, Nanchong 637002, China; \\
        \and
             Key Laboratory for Research in Galaxies and Cosmology, Shanghai Astronomical Observatory, Chinese Academy of Sciences,80 Nandan Road, Shanghai 200030, China. \\
             \and
              School of Astronomy and Space Science, University of Chinese Academy of Sciences, No. 19A, Yuquan Road, Beijing 100049, China. \\
             {\it e-mail: jzhong@shao.ac.cn \& lijing@bao.ac.cn}
}

\abstract{ COIN-Gaia 13 is a newly discovered open cluster revealed by Gaia DR2 data. It is a nearby open cluster with a distance of about 513 pc. Combined with the five-dimensional astrometric data of Gaia EDR3 with higher accuracy, we use the membership assignment algorithm (pyUPMASK) to determine the membership of COIN-Gaia 13 in a large extended spatial region. The cluster has found 478 candidate members. After obtaining reliable cluster members, we further study its basic properties and spatial distribution. Our results show that there is an obvious extended structure of the cluster in the X-Y plane. This elongated structure is distributed along the spiral arm, and the whole length is about 270 pc.  The cluster age is 250 Myr, the total mass is about 439 M$_\odot$, and the tidal radius of the cluster is about 11 pc. Since more than half of member stars (352 stars) are located outside twice of the tidal radius, it is suspected that this cluster is undergoing the dynamic dissolution process. Furthermore, the spatial distribution and kinematic analysis indicate that the extended structure in COIN-Gaia 13 is more likely to be caused by the differential rotation of the Galaxy.
}
  \authorrunning{Bai et al}            
  \titlerunning{The extended structure of COIN-Gaia 13}  

  \maketitle
%

\section{Introduction}       
Most open star clusters are located on the spiral arm, composed of young stars in the Galactic thin disk (compared with globular clusters), and are in continuous formation \citep{2016EAS....80...73M}. It is generally believed that stars in clusters are born in the same high-density molecular cloud region and have similar properties such as age, kinematic and chemical characteristics, and so on. Then, star clusters will be affected by many factors in the process of dynamic evolution which lead to mass loss, such as two body relaxation\citep{1987degc.book.....S}, disk impact\citep{1987gady.book.....B} and tidal forces\citep{1958ApJ...127...17S,2009gcgg.book..375G} (either by the Galactic gravitational potential or giant molecular clouds or other star clusters). Finally, with the expansion and disintegration of star clusters，most member stars are dispersed as field stars. Therefore, open clusters are important tools for studying the formation and evolution of our Galaxy (especially the Galactic disk). 

In general, the first step of studying open cluster is to determine the member stars. The reliability of cluster member identification greatly depends on the accuracy of observational data. The {\it Gaia} mission \footnote{(\url{https://www.cosmos.esa.int/gaia})} provides precise five astrometric parameters ($l$, $b$, $\varpi$, $\mu_{\alpha}^*$, $\mu_{\delta}$) and three band photometry ($G$, $G_{BP}$ and $G_{RP}$) for more than one billion stars \citep{2018A&A...616A...2L}. This allows us to investigate a large number of open clusters with unprecedented accuracy. After the data release of Gaia DR2 \citep{2018A&A...616A...4E}, the study of open star clusters has ushered in a new era: on the one hand, a large number of member stars in reported star clusters were re-identified\citep{2018A&A...618A..93C}; on the other hand, more and more newly discovered star clusters are revealed and reported.  \citet{2018A&A...618A..93C,2020A&A...633A..99C,2020A&A...640A...1C} used the UPMASK algorithm to determine the cluster members and provided an updated cluster catalog including previously known clusters and newly discovered clusters. 

Due to the improvement of data accuracy by Gaia, it is possible to detect the low-density structure of clusters in the position space. For example, the newly discovered young extended structure of the Double Cluster $h$ and $\chi$ Persei has a scale of about six to eight times the core radii \citep{2019A&A...624A..34Z}. Furthermore, \citet{2019A&A...624A..34Z} also report the discovery of filamentary substructures extending to about 200 pc away from the Double Cluster. Similar discoveries of extended structures are the star relic filaments with dozens of pc scales in the Orion star-forming region \citep{2019MNRAS.489.4418J}. Using the Gaia data, there are many studies focused on the tidal tails of nearby old clusters:  \citet{2019A&A...627A...4R} searched the tidal tail of Praesepe (NGC 2632); \citet{2020ApJ...889...99Z} study the tidal structure of Blanco 1; the extended tidal structure of the Hyades were reported by two works\citep{2019A&A...621L...3M,2019A&A...621L...2R}. 
By exploring the extended region of star clusters, we now know that in addition to the core with approximate symmetric distribution, there are low-density elongated or asymmetric structures in the outskirt of many star clusters. These discovery are helpful for us to better understand the influence of formation environment on stars and clusters, as well as initial condition of cluster dynamic evolution.

COIN-Gaia 13 is a newly discovered young open clusters by Gaia DR2 data \citep[hereafter CG19]{2019A&A...624A.126C}. We attempt to explore the spatial structure up to the outmost reach of this nearby cluster $(~500pc)$ and investigate its dynamic evolution. In this paper, we search for members of COIN-Gaia 13 in a large region using the latest Gaia EDR3 data to study the properties of the cluster in more detail. In section~\ref{data}, we mainly introduce our selection criteria and member determination methods. In section~\ref{res}, we provide the identification results of member stars and the distribution properties of member stars in different parameter space. Finally, in Section~\ref{Summary}, we briefly summarize and discuss our results.

\section{Data and method}
\label{data}
\subsection{Data process}
\label{Data process}

\begin{figure}
\centering
\includegraphics[scale=0.5]{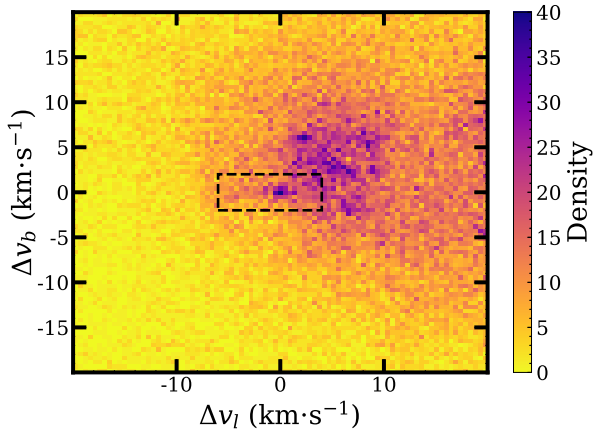}
\caption{Velocity distribution of the stars in the tangential velocity plane ($\Delta v_{l},\Delta v_{b} $ ). The zero-point is the average motion value of COIN-Gaia 13 in CG19. The box indicates the selection window that might belong to the cluster.}
\label{fig1}
\end{figure}

As the early stage of Gaia's third data release, Gaia EDR3 \citep{2021A&A...649A...1G} provides us astrometric and photometric parameters of 1.5 billion sources with higher accuracy than Gaia DR2. In this paper, we use Python astroquery package \citep{2013A&A...558A..33A,2018AJ....156..123A} to retrieve the Gaia EDR3 data source samples we need from Gaia science archive.
In the cluster catalog provided by \citet{2019A&A...624A.126C}, the galactic coordinates (l,b) of COIN-Gaia 13 is (167.459$^{\circ}$, 4.776$^{\circ}$), the average proper motion ($\mu \alpha$,$\mu \delta$) is (-3.83, -1.66) mas$\cdot$yr$^{-1}$, and the average parallax is 1.93 mas.Based on the previous research \citep{2019AJ....158..122K}, it is noted that our cluster is mainly distributed along the Galactic plane(the distribution along the galactic longitude is not symmetrical about the center), so we only include stars with Galactic longitude coordinate between 135 to 185 degrees and the Galactic latitude between -2  to 12 degrees as the candidate sources, which is large enough to cover the whole area of the cluster. Then, in order to make sure the reliability of astrometric data, we select stars whose G-band magnitude are brighter than 18 magnitudes and the renormalized unit weight error (ruwe) is less than 1.4. We further limit the parallax of stars between 1.6 mas and 2.4 mas to exclude field stars in the background. In order to better perform the clustering algorithm in the kinematic space, we convert the proper motion angular velocity (mas$\cdot$yr$^{-1}$) into tangential velocity (km$\cdot$s$^{-1}$) using the formula as  V=4.75$\mu$d, where $\mu$ is the proper motion and d is the distance (as the inverse of parallax), and further subtract the average tangential velocity which provided by \citet{2019A&A...624A.126C}. In Figure~\ref{fig1}, we present the relative distribution of tangential velocity of candidate sources, and notice that an obvious over-density pattern extends along the x-axis. Considering these co-moving stars in the over-density region are more likely to be member stars, we cut out a window between $-6$ to $+4$ km$\cdot$s$^{-1}$ in $\Delta v_{l}$ and $-2$ to $+2$ km$\cdot$s$^{-1}$ in $\Delta v_{b}$ to select candidate sources. After these selection criteria, we finally obtained a total of 3207 candidates (labeled sample 1) to perform the membership identification.

\subsection{Method}
\label{method}
In this paper, we mainly use the pyUPMASK algorithm \citep{2021A&A...650A.109P} to perform the membership determination. The pyUPMASK algorithm is an open source software package compiled by Python language following the development principle of UPMASK \citep{2014A&A...561A..57K}, which is the short name of Unsupervised Photometric Membership Assignment in Stellar Clusters. It is a member star determination method developed to process photometric data at first, however, it was later widely used in the determination of member stars based on astrometric parameters \citep{2018A&A...618A..93C,2020A&A...633A..99C,2020A&A...640A...1C}.  In order to obtain the membership probability of cluster members more effectively and quickly, pyUPMASK improves the clustering method, member determination step and membership probability assignment method in its determination process. Furthermore, in case of the situation with serious field stars contamination, a further field star excluding process in spatial space is added after the member determination process. These improvements not only make the pyUPMASK algorithm more robust for cluster membership determination, but also greatly reduced the computer time and makes it more effective to process a large number of sources. 

The key assumptions of UPMASK include: (I) cluster members have the similar properties (for example, they have the clustering distribution both in proper motion and parallax); (II) their spatial distribution is more crowded than a random uniform distribution. The main steps of this method include: (I) Use the K-means clustering method to determine the clumps in three-dimensional astrometric space ($\mu_l $, $\mu_b $, $\varpi$);(II) According to the estimation results of kernel density, determine whether these small clumps have the clustering distribution in all parameter space (compared with a random uniform distribution). If so, it is preliminarily determined that the star in its small clump is a cluster member with similar spatial and kinematic properties.  

We apply the pyUPMASK algorithm to the preliminarily selected sample sources, which are labeled as sample 1 (see more details in section~\ref{Data process}), and then obtain the cluster member probability of each star.  In the process of determining membership probabilities, this algorithm will mark the samples as members or non-members.The pyUPMASK mainly adopts a k-means clustering algorithm, which is a method of data clustering according to the distance from each center point. In order to reduce the inconsistency of clustering results due to the randomness of the initial center point assignment, the pyUPMASK uses the same input data to perform the multiple allocation and decision process. Finally, each star has multiple decision results, and the member probability is the frequency of stars marked as members (P = n / N, where N is the repeating times, and n is the number of times when a single star is determined as a member). According to the probability function adopted by pyUPMASK, field stars are often assigned smaller member probability values.We investigate the properties of stars with P $<$ 0.5 and found that they present a uniform distribution in position space and have no obvious clumping structure in the proper motion space. On the other hand, for stars with P $>$ 0.5, the clear main sequence distribution in the color-magnitude diagram (Figure~\ref{fig5}) illustrates the low contamination rate of our member candidates. The criteria probability of 0.5 is also consistent with the criteria of other works \citep[e.g.,][]{2019A&A...626A..17C,2019Ap&SS.364..152Y, 2021Ap&SS.366...68A}. Therefore, we select stars whose membership probabilities greater than 0.5 as cluster members. Finally, a total of 478 candidate members of COIN-Gaia 13 were obtained.

It is noted that the contamination and completeness of the sample of cluster members which were identified by the pyUPMASK algorithm have good performance.In order to estimate these two parameters with Gaia data, we simulated ten mock samples composed of 1200 member stars (similar to M67 parameter distribution) and 1200 background field stars ( with random and uniform distribution in the same phase space as members). And then use the pyUPMASK to perform the membership determination. The results of ten mock samples show that the average completeness of the cluster is about 97\%, and the average contaminate rate is 4\%. Although the completeness rate slightly decreases with the magnitude increasing, the rate can go to 95\% even in the G=18 mag.

\section{Results and Discussion}
\label{res}
\subsection{Cluster members}
Our research is a further detailed study of COIN-Gaia 13 reported by CG19. The spatial distribution of sample 1 with its membership probability is shown in figure~\ref{fig2}, in which purple dots represent member stars with high probabilities and present an elongated distribution. In figure~\ref{fig3}, we present the histogram of membership probability of all sources determined by the pyUPMASK.  It is noted that stars with low membership probability are more likely to be field stars, so we only select stars with P $>$ 0.5 as member stars to ensure that the spatial structure traced by member stars are more reliable.  Although the criteria of membership probability we selected is 0.5, the membership probability of most member stars in our sample is greater than 0.8, which also means that the field star contamination in our sample is very small. Since we select an extended large region to perform the membership identification, there are more member candidates in our sample (478 members)  than in CG19 (171 members with P $>$ 0.5).

\subsection{Spatial distribution}
\begin{figure}
\centering
\includegraphics[scale=0.43]{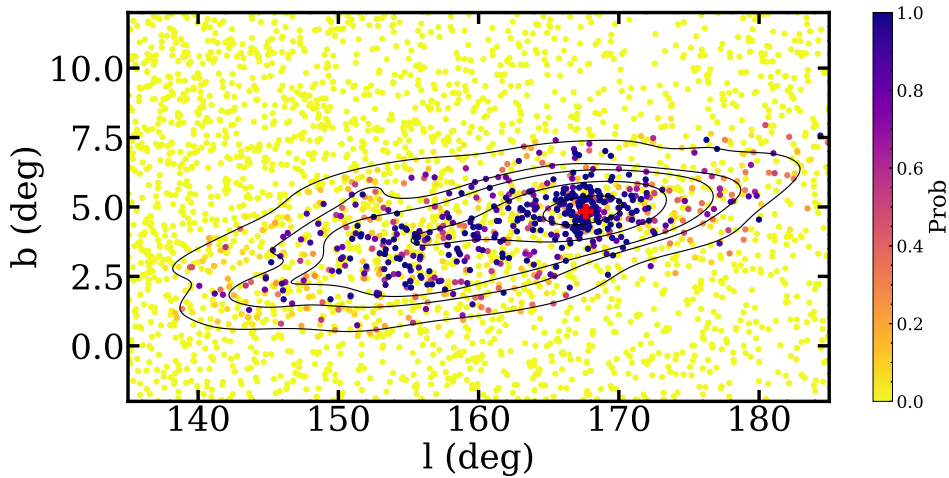}
\caption{Spatial distribution of sample 1 in the Galactic coordinate. It is noted that a large number of member stars are located outside of the cluster symmetrical core region, which shows an extended elongated structure along the Galactic meridian direction.The red cross represent the cluster center with coordinate ($l$=167.645$^{\circ}$,~$b$= 4.877$^{\circ}$).}
\label{fig2}
\end{figure}

\begin{figure}
\centering
\includegraphics[scale=0.4]{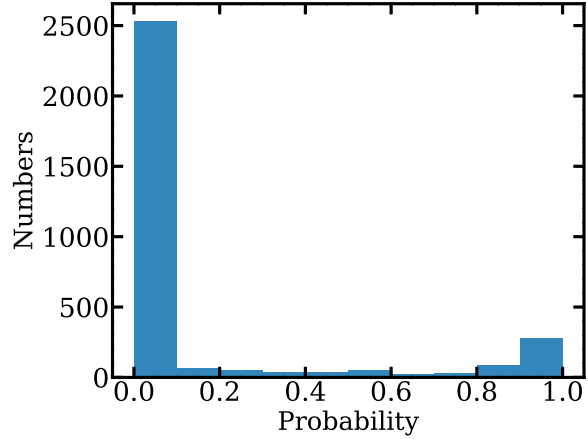}
\caption{Histogram of membership probability of COIN-Gaia 13. The distribution shows that the probability determined by the pyUPMASK can well distinguish the member stars and field stars.}
\label{fig3}
\end{figure}

The cluster center is defined as the location of the highest density area of cluster members. In order to obtain the maximum central density of star clusters, we use the kernel density estimation function to estimate the distribution of members in the cluster region. After the calculation of two-dimensional Gaussian kernel density, the probability density distribution of star clusters is shown in figure~\ref{fig2}. The center coordinates ($l$=167.645$^{\circ}$,~$b$= 4.877$^{\circ}$) of COIN-Gaia 13, which represent the highest density location of the cluster, is similar with the reference values ($l$=167.459$^{\circ}$,~$b$= 4.776$^{\circ}$) in CG19. In figure~\ref{fig2}, we use the red cross to represent the cluster center. 
The discovery of elongated tail structure is our major result. Figure~\ref{fig2} shows that the extended direction of this tail is along with the Galactic longitude. Compare with the elongated structure along with the line of sight, the extended structure along with the projection direction is more likely to be a real structure. This is because the  position precision is much higher than the parallax precision, even in the Gaia data.

\begin{figure}
\centering
\includegraphics[scale=0.20]{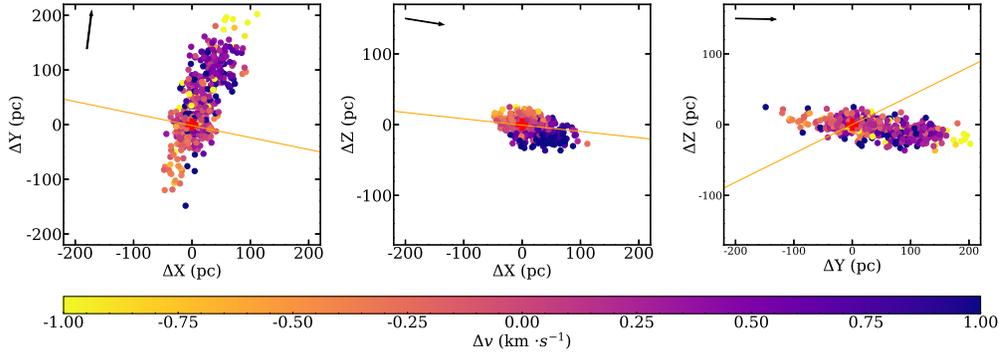}
\caption{Relative spatial distribution of member candidates after correcting the position coordinates of the cluster center. The corresponding position center of COIN-Gaia 13 is (X$_c$, Y$_c$, Z$_c$) = ( -8621.0, +109.3, +65.3) pc. We use a yellow line to represent the line of sight direction. The major elongated structure is approximately perpendicular to the line of sight direction, which further proves the authenticity of this structure. In each spatial projection panel, we use colors to represent relative velocities which direction is along with the major motion direction, and the mean projected velocity of core members is as its zero point. The black arrow points out the major motion direction.}
\label{fig4}
\end{figure}

Furthermore, we calculate the Galactocentric coordinate $(X,Y,Z)$ for all determined member stars. The coordinate system takes the Galactic center as the coordinate origin, the direction from the sun to the Galactic center is the positive direction of the X axis, the Y axis points to the rotation direction of the Galactic disk, and the Z axis points to the North Galactic pole. The solar location of Galactocentric coordinate is adopted as (-8122, 0 ,20.8) pc ~\citep{2019A&A...622A.205K}. Figure~\ref{fig4} shows the relative spatial distribution after correcting the position coordinates of the cluster center. The corresponding position center of COIN-Gaia 13 is (X$_c$, Y$_c$, Z$_c$) = ( -8621.0, +109.3, +65.3) pc. The major elongated structure in the Galactocentric coordinate is along with the Y axis, which is approximately perpendicular to the line of sight direction. Since the spatial distribution of members is based on the astrometric data with high precision, the extended structure in the Y direction has a high probability to be a real structure instead of a spurious structure which due to the observational errors. Considering the average distance of COIN-Gaia 13 is 513 pc (see next subsection), the physical scale of the elongated structure in the Y direction is about 270 pc. For comparison, the distribution lengths of cluster members in X and Z directions are about 130 pc and 50 pc respectively. This also shows that the cluster is mainly distributed in the Galactic plane, while the dispersion in the Z direction is very small.

\subsection{Isochrone fitting}
\label{iso}
\begin{figure}
\centering
\includegraphics[scale=0.5]{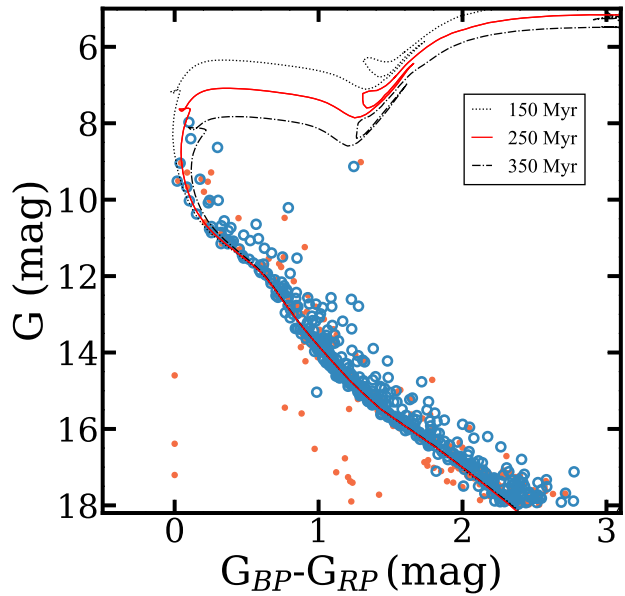}
\caption{Colour-magnitude diagram of member candidates in COIN-Gaia 13. Blue dots and orange dots represent 478 member candidates in this work and 311 members in CG19, respectively.PARSEC isochrones of 150, 250, and 350 Myr with solar metallicity and the same extinction are over-plotted.
The PARSEC isochrone with the age of 250 Myr provide the best fitting results.}
\label{fig5}
\end{figure}

The color magnitude diagram (CMD) is an important tool to estimate the fundamental parameters of star clusters. After fitting the observed color-magnitude distribution of member stars by theoretical isochrone, we can obtain important cluster parameters with age, distance modulus and reddening. Figure~\ref{fig5} shows the cluster members distribution in CMD. For comparison, we also plot the member candidates distribution provided by CG19 with orange dots. It is noted that our membership determination provide more reliable result than previous study. 

In our work, we use the theoretical isochrone from PARSEC \citep {2012MNRAS.427..127B,2014MNRAS.444.2525C,2017ApJ...835...77M} to perform the CMD fitting through visual inspection. Since there is no abundance parameter of the COIN-Gaia 13, we adopt the solar metallicity [Z/X]$_{\odot}$ = 0.0207 to perform the isochrone fitting. The best-fitting result is shown on figure~\ref{fig5},  in which the age is 250 $\pm$ 100 Myr (logt = 8.4), the distance modulus  is 8.9 mag, and the extinction in G-band is  0.32 mag. The derived distance of COIN-Gaia 13 by isochrone fitting is 510 pc. We also derived the average parallax of cluster members plx=1.95 $\pm$ 0.08 mas, which correspond to the distance of 513 pc. 

\subsection{kinematic properties}
\begin{figure}
\centering
\includegraphics[scale=0.5]{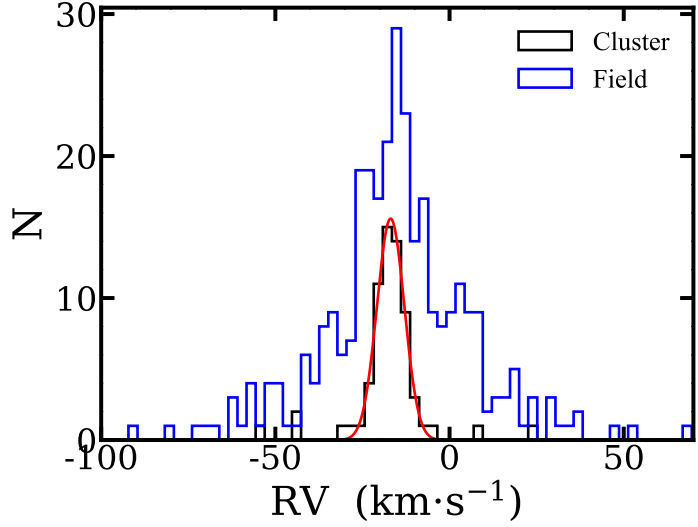}
\caption{Histogram of line of sight velocity distribution of 66 members. The average line of sight velocity of cluster is -16.9 $\pm$ 4.0 km$\cdot$s$^{-1}$. As a comparison, the dispersion of non-member stars in the background field is 18.3 km$\cdot$s$^{-1}$. }
\label{fig6}
\end{figure}

Cross-match the 460  cluster members with the LAMOST DR8, LAMOST-LRS data \citep{2012RAA....12..723Z} provide us with  the measured line of sight velocities of \textbf{82} member stars, of which 66 have a measurement error of less than 8 km$\cdot$s$^{-1}$. Although the number of RV is relatively small, there is still an obvious peak in the histogram distribution of the line of sight velocity (Figure~\ref{fig6}). By fitting a Gaussian distribution, the average line of sight velocity is   -16.9 km$\cdot$s$^{-1}$, with the dispersion of 4.0 km$\cdot$s$^{-1}$. 

\begin{figure}
\centering
\includegraphics[scale=0.25]{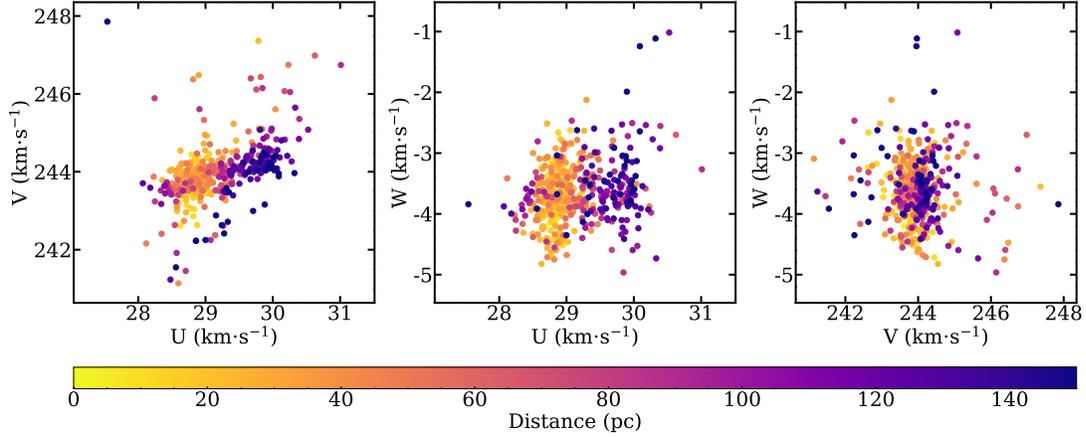}
\caption{Distribution of UVW components of motion for all members.  The asymmetric pattern in the velocity space (U-V panel) happens to correspond to the extended structure in the spatial distribution ( X-Y panel), clearly indicating the extended structure is the result of cluster dynamic evolution. In addition, we use colors to represent the relative distance of members from the cluster center and note that the velocity distribution of members has an obvious correlation with the distance, which also indicates that the elongated structure is not formed in the original.}
\label{fig7}
\end{figure}

Due to the insufficient line of sight velocity of cluster members provided by the LAMOST, we assign each member star a line of sight velocity, and then calculate their (U,V,W) space velocities. We note that the extended spatial structure over $\pm$ 20 degrees on the sky will lead to an offset of line of sight velocity from the core region to the outer region. We then assign the line of sight velocity of each star as V$_{r}$=V*cos$\theta$, where V is the mean line of sight  velocity ( -16.9 km$\cdot$s$^{-1}$) and $\theta$ is the position angle from the cluster center to the star. The (U,V,W) velocity distribution of members is shown in Figure~\ref{fig7}, while the average velocity of(U, V, W)= (+28.86，+243.83，-3.82) km$\cdot$s$^{-1}$. It is noted that the cluster shows an correlation in the U-V and U-W velocity diagrams, which happens to correspond to the extended structure in the spatial distribution. Furthermore, we use color to represent the relative distance between the cluster center to the three-dimensional location of members, and note that the velocity of U / V has a strong correlation with the relative distance: the farther away from the center, the greater difference of the U / V velocity. In addition, in Figure~\ref{fig4}, we also use colors to represent relative velocities which direction is along with the major motion direction, and the mean projected velocity of core members is as its zero point. It is obvious that the velocity of cluster members has a significant correlation with the spatial distribution, especially in the Y direction, which also indicates that the elongated structure is not initially formed with the cluster. This is because if the elongated structure is primordial, then all member stars should be shared with similar velocity and independent of the spatial location. In contrast, if member stars are originated from the cluster center, the correlation between velocity and position is a natural evolution result：the higher velocity of stars, the more distant the spread location.  

\begin{figure}
\centering
\includegraphics[scale=0.3]{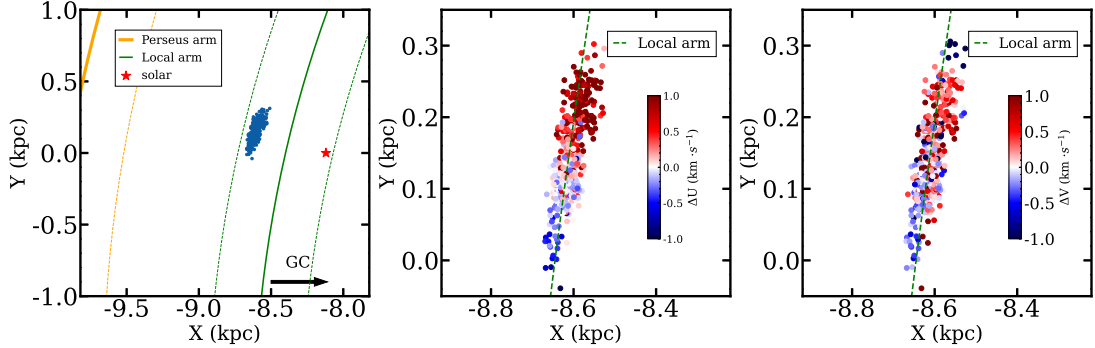}
\caption{ Spatial distribution of COIN-Gaia 13 in the Galactocentric X-Y plane. The left panel shows the relative positional distribution between the cluster and nearby spiral arms, while red stars represent our solar location. In the middle and right panels, the relative velocities of member stars in U and V directions   ( with the mean velocity of the cluster center as zero point) are represented by colors, respectively.  It would appear that the relative velocity has an obvious correlation with the local arm direction both in the U and V velocity components, which further suggests that the elongated structure in COIN-Gaia 13 is more likely to be caused by the differential rotation at different positions relative to the Galactic center. }
\label{fig8}
\end{figure}

The evolved elongated structure of star clusters may be caused by the tidal force of surrounding gravity sources (like giant molecular clouds) or differential rotation in the Galactic disk. We further study the spatial distribution and velocity distribution of member stars compared with the Galactic spiral arm. The location of spiral arms is obtained by \citet{2014ApJ...783..130R}, which mainly uses the VLBI observation of maser sources to trace the arm structure distribution. The distance of COIN-Gaia 13 is 513 pc, which is located in the nearby region of the local arm. Figure~\ref{fig8} present the relative position distribution between COIN-Gaia 13 and the nearby spiral arms (left panel), and the relative velocity distribution (with the mean velocity of the cluster center as zero point)  in the X-Y Galactocentric coordinate diagram ( middle and right panel). It is noted that the spatial distribution and the relative velocity offset (especially in the U direction component) of the elongated structure are almost along with the local arm. Considering the relative velocity offset is on the order of 1 $\sim$ km$\cdot$s$^{-1}$, which is enough to stretch the cluster to more than 200 pc in a time scale of 200 Myr. Therefore, we suggest that the elongated structure in COIN-Gaia 13 is more likely to be caused by the differential rotation at different positions relative to the Galactic center.       

\subsection{Mass function}
\begin{figure}
\centering
\includegraphics[scale=0.4]{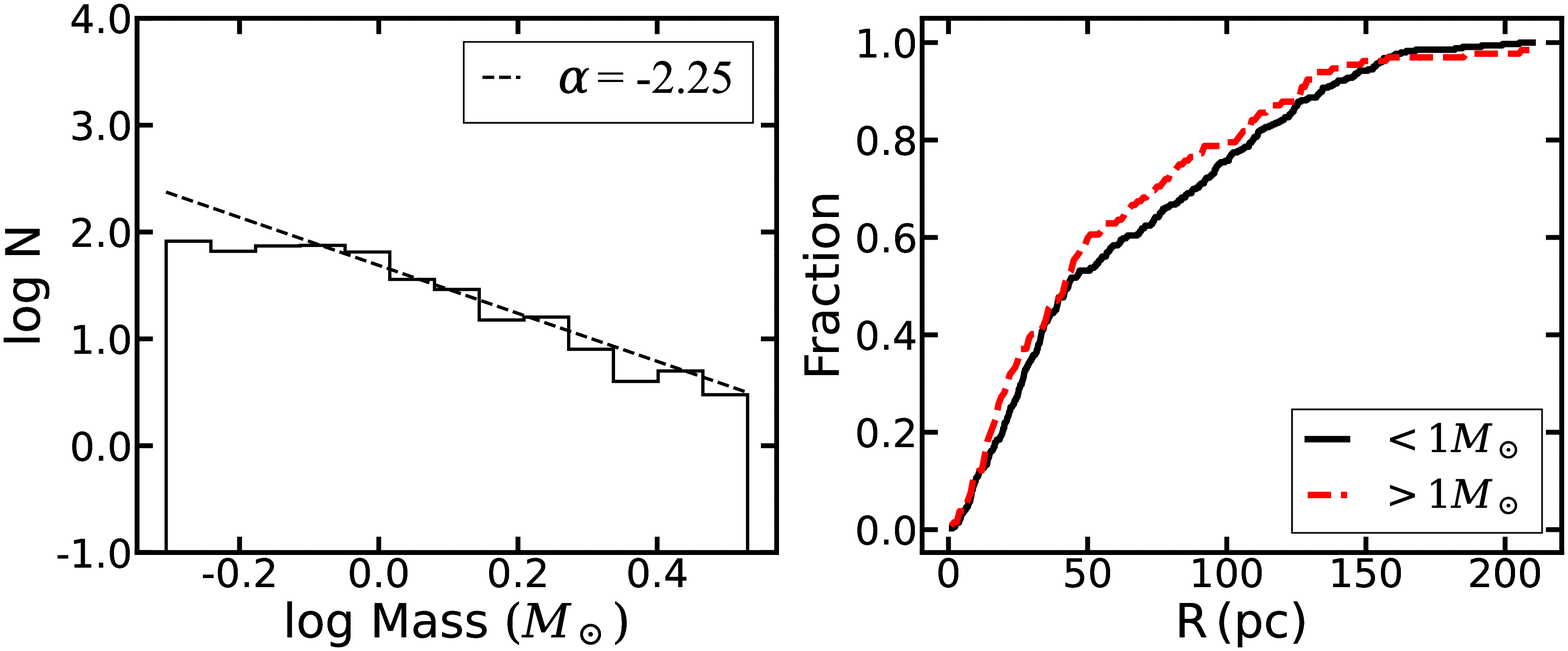}
\caption{Diagram of mass function distribution (left panel) and cumulative fraction as a function of radial distance from cluster center (right panel). The dashed line in the left panel presents a power-law distribution with slope $\alpha$=-2.25. In the right panel, the similar cumulative mass distribution between the high mass group and the low mass group shows that there is no obvious mass segregation in the cluster COIN-Gaia 13, which indicates that the extended structure is unlikely to be formed by the tidal force of surrounding gravity sources. }
\label{fig9}
\end{figure}

The open cluster contains hundreds of stars with the same age and chemical composition but different masses. In order to study the mass distribution of member stars, it is an efficient method to perform the statistics of the mass function. According to the evaluation results of completeness by the pyUPMASK(See Section~\ref{method}), the completeness decreases slowly with the increase of magnitude, which indicates that the slope of mass function estimated by the star count is reliable since the influence of completeness effect on the determination of mass function is not significant. We first perform the isochrone fitting of all member stars, and then estimate their stellar mass using the PARSEC isochrone with 250 Myr. Finally, we obtain the present distribution of mass function (PDMF) of the cluster COIN-Gaia 13, which is shown in figure~\ref{fig9}.  The dotted line in the left panel of figure~\ref{fig9} is represented the fitting of the power-law function with slope $\alpha = -2.25\pm0.14$.  We only include member stars with G-band magnitude brighter than 18 mag ( mass greater than 0.5 M$_{\odot}$), the total mass of the cluster is 439 M$_{\odot}$. However, considering the incomplete membership determination results and the exclusion of binary stars, we note that the mass function is still underestimated. 

The tidal radius of the cluster can be estimated according to the cluster total mass, which is given by  \citet{1998MNRAS.299..955P} with the formula:
\begin {equation}
r_ t=\left(\frac{GM_C}{2(A-B)^2}\right)^\frac{1}{3}
\end{equation}  
where $G=4.3\times 10^{-3}$pc$\cdot$ M$_\odot^{-1}$(km$\cdot$ s$^{-1})^{2}$ is the gravitational constant,  $A=15.3\pm0.4$ ~km$\cdot$ s$^{-1}\cdot$ kpc$^{-1}$ and $B=-11.9\pm0.4$ ~km$\cdot$ s$^{-1}\cdot$ kpc$^{-1}$ are Oort constants related to the cluster orbital position. According to the cluster total mass $M_ C$ = 439 M$_{\odot}$, the tidal radius of COIN-Gaia 13 is estimated to be 11 pc. Within the twice of the tidal radius, there are 126 members, while more members (352) are located beyond this region. 

To speculate on the possible origin of the extended structure, we study the mass segregation of cluster members. In the dynamical evolution of a cluster, the higher mass stars would give up their energy to the lower mass stars and then drop down to the cluster core region\citep{1996Ap&SS.235...93Z}. If the cluster is located in the gradient of gravitational potential caused by a nearby gravity source (such as the giant molecular cloud), stars in the outer region (more likely to be low mass stars) would obtain more energy and move to the more distant orbit. This means that if there is a gravity source close to the cluster, the mass segregation effect will be more significant.  As shown in the right panel of figure ~\ref{fig9}, the cumulative mass distribution as a function of radial distance in high mass and low mass groups presents similar distribution and does not increase significantly from the inside to the outside. It appears that there is no obvious mass segregation in the cluster COIN-Gaia 13 and further indicates that the extended structure is unlikely to be formed by the tidal force of the nearby gravity source.

\section{Summary}
\label{Summary}
Using the five-dimensional data provided by Gaia EDR3 ( spatial coordinates, proper motions and  parallax), we studied the COIN-Gaia 13 in detail. For the samples located in a large extended spatial region , we use the membership assignment algorithm pyUPMASK to assign the membership probability to each star. We selected stars with membership probability greater than 0.5 as reliable results and finally obtained 478 cluster member candidates of COIN-Gaia 13. 

We derived the fundamental parameters of the cluster COIN-Gaia 13, including the Galactic coordinate of the cluster center location (167.645$^{\circ}$, 4.837$^{\circ}$), the average parallax 1.95 mas, the average proper motion ($\mu l$, $\mu b$) =(-0.696, -4.184) mas$\cdot$yr$^{-1}$,  and the average line of sight velocity -16.9 $\pm$ 4.0 ~km$\cdot$s$^{-1}$.  After converting the observed coordinates to the Galactocentric coordinates, the spatial position is ($X_c $, $Y_c $, $Z_c $) = ( -8621.0, + 109.3, +65.3) pc, and the spatial motion is ($U_c$，$V_c$，$W_c$)=（+28.86，+243.83，-3.82）km$\cdot$s$^{-1}$. By fitting the CMD distribution of high probability members with PARSEC isochrone, the cluster age and total mass are estimated as 250 Myr and 439$M_\odot$ respectively. 

The major result of our work is the discovery of an extended structure of the cluster COIN-Gaia 13, which is a elongated structure with length about 270 pc. Further analysis indicates that the extended structure is more likely to be caused by the differential rotation along with the local arm direction. 

{\bf Acknowledgments}
We are very grateful to the referee for helpful suggestions which have improved the paper significantly.
This work is supported by National Key R\&D Program of China No. 2019YFA0405501. Li Chen acknowledges the support from the National Natural Science Foundation of China (NSFC) through the grants  12090040 and 12090042. Jing Zhong would like to acknowledge the National Science Foundation of China (NSFC) under grants 12073060, and the Youth Innovation Promotion Association CAS. Jing Li would like to acknowledge the Innovation Team Funds of China West Normal University and the Sichuan Youth Science and Technology Innovation Research Team (21CXTD0038).

This work has made use of data from the European Space Agency (ESA) mission GAIA (\url{https://www.cosmos.esa.int/gaia}), processed by the GAIA Data Processing and Analysis Consortium (DPAC,\url{https://www.cosmos.esa.int/web/gaia/dpac/consortium}). Funding for the DPAC has been provided by national institutions, in particular the institutions participating in the GAIA Multilateral Agreement.

\end{CJK*}
\end{document}